\documentclass[pra,twocolumn,showpacs,
superscriptaddress,unsortedaddress]{revtex4}

\usepackage{latexsym}
\usepackage{amssymb}
\usepackage{amsmath}
\usepackage{exscale}
\usepackage{bm}
\usepackage{graphicx}
\usepackage{psfrag}

\newcommand{\rb}{\mathbf{r}}
\newcommand{\di}{\mathrm{d}}
\newcommand{\db}{\mathbf{d}}
\newcommand{\bc}{\begin{center}}
\newcommand{\ec}{\end{center}}
\newcommand{\ints}{\int_0^{\prime\infty}}
\newcommand{\intss}{\int_0^{\prime\prime\infty}}
\newcommand{\Ints}{\ints}
\newcommand{\Intss}{\intss}

\newcommand{\sumprime}{\hspace{-.5ex}{^{\textstyle'}}}

\begin{document}

\title{%
Generation of long-living entanglement between two
separate
atoms 
}

\author{\"Ozg\"ur \c{C}ak\i r}

\affiliation{%
Department of Physics,
Bilkent University,
06533 Bilkent, Ankara, Turkey}

\author{Ho Trung Dung}

\affiliation{%
Institute of Physics, National Center
for Sciences and Technology, 1 Mac Dinh Chi Street,
District 1, Ho Chi Minh city, Vietnam}

\author{Ludwig Kn\"{o}ll}

\author{Dirk-Gunnar Welsch}

\affiliation{%
Theoretisch-Physikalisches Institut,
Friedrich-Schiller-Universit\"{a}t Jena,
Max-Wien-Platz 1, 07743 Jena, Germany}

\date{\today}

\begin{abstract}
A scheme for non-conditional generation of long-living 
maximally entangled states between two spatially well
separated atoms is proposed.
In the scheme, $\Lambda$-type atoms pass a 
resonator-like equipment of dispersing and absorbing macroscopic 
bodies giving rise to body-assisted electromagnetic field resonances
of well-defined heights and widths.
Strong atom-field coupling is combined with weak atom-field coupling
to realize entanglement transfer from the dipole-allowed
transitions to the dipole-forbidden transitions, thereby
the entanglement being preserved when the atoms depart 
from the bodies and from each other.
The theory is applied to the case of the atoms passing
by a microsphere.
\end{abstract}

\maketitle

\section{Introduction}
Generation of entanglement in atomic systems has been a subject
of intense theoretical and experimental study motivated by both the
fundamental issue and potential applications in quantum information
processing. In this context, the realization of easily
controllable long-living entangled states of spatially well 
separated atoms has been one of the crucial and challenging problems.
A number of methods of entanglement preparation between atoms
have been considered such as
the use of quantum-correlated light fields interacting with
separate atoms, thereby transferring their entanglement to the
atoms \cite{polzik99,lukin00,kuzmich00,molmer00,lloyd01,kraus04},  
appropriate measurements on the light in multi-atom--light interaction
processes, thereby conditionally projecting the atoms in entangled
states \cite{cabrillo99,bose99,duan00,julsgaard01,%
duan01,duan02,gilchrist02,feng03,duan03,simon03,browne03}, and 
the technique of quantum reservoir engineering in a cascaded
cavity-QED setting \cite{clark03}.

Photon exchange between two atoms is one of the
simplest processes to entangle two atoms in
a common electromagnetic field.
The effect, which is very weak in free space,
can be enhanced significantly when the atoms are in a cavity
\cite{hagley97,rauschenbeutel00,osnaghi01}.
Usually attempts are made to minimize the effect 
of spontaneous emission. Quite counterintuitively,
in certain situations one can take advantage of
the spontaneous emission for entanglement generation
\cite{ho02,Basharov02,Ficek03,Kastel04}. Consider, for example, 
two two-level atoms located in free space
with one of them being initially excited. 
This product state is a superposition of 
a symmetric (superradiant) state and an antisymmetric 
(subradiant) state. If the two atoms are separated by 
distances much smaller than the wavelength, the symmetric state
decays must faster than the antisymmetric one, leaving the
system in a mixture of the ground state and 
the entangled antisymmetric state. 

The scheme also works at distances much larger than the
wavelength, if a resonator-like equipment is used which
sufficiently enhances the atom-field coupling,
thereby ensuring that a photon emitted in the process of 
resonant photon exchange, which is mediated by real photon
emission and absorption, is accessible to the two atoms.
This condition can be satisfied, for instance,
when the atoms pass by a dielectric microsphere
at diametrically opposite positions \cite{ho02}.
If the distance of the atoms from the surface of the
sphere becomes sufficiently small, then the
excitation of surface-guided (SG) and whispering
gallery (WG) waves can give rise to
strong collective effects, which are necessarily
required to generate substantial entanglement.
Needless to say that other than spherically symmetric
bodies can also be used to realize a noticeable mutual coupling
of the atoms.    

A drawback of the use of two-level-type atoms
is that the entanglement is transient. In particular,
when two atoms that have become entangled between
each other near a body such as a microsphere move away from it
(and from each other), then they undergo ordinary spontaneous
emission (in free space), which destroys the quantum coherence.
Preservation of the atomic entanglement over long distances
between the atoms is therefore not possible in this way.

The contradicting effects of entanglement creation and
destruction typical of two-level atoms can be combined in a
more refined scheme 
involving two three-level atoms of $\Lambda$ type each
(Fig.~\ref{fig1}), where the two lower lying states
$|1\rangle$ and $|2\rangle$ such as the ground state and
a metastable state or two metastable states represent the qubits
that are desired to be entangled with each other \cite{can03}.
Whereas the transition $|1\rangle$ $\!\leftrightarrow$ 
$|3\rangle$ is strongly coupled to the field, the transition
$|2\rangle$ $\!\leftrightarrow$ $|3\rangle$ is 
only weakly coupled to the field. Each atom is initially
in the state $|1\rangle$, while the field
is prepared in a single-photon state. Let us assume that
due to Rabi oscillations the state $|3\rangle$ of one of the 
two atoms, we do not know which one, is populated. Irreversible
decay to the state $|2\rangle$ is then accompanied with
an entanglement transfer forming a (quasi-)stationary
entangled state between the two atoms with respect to the states
$|1\rangle$ and $|2\rangle$. Its lifetime is limited only by the
lifetime of the metastable states, 
and the degree of entanglement
achievable can approach 100\% in principle. 
Moreover, the scheme is 
non-conditional
and realizable by means of current experimental techniques.

In fact, the model Hamiltonian used in Ref.~\cite{can03} is
based on a Dicke-type system and does not allow for
atoms that are spatially well separated from each other, 
with the interatomic distance being much larger than the
characteristic wavelengths.
However, for many applications in quantum information
processing or for testing Bell's inequalities, large
interatomic distances and thus the possibility of
individual manipulation of the atoms are necessary prerequisites.    
The aim of the present paper is to close this loophole,
by considering two spatially well separated $\Lambda$-type
three-level atoms appropriately positioned with
respect to macroscopic bodies,
so that the two key ingredients --
enhanced atom-field coupling and sharp field resonances
can be realized. Note that the second ingredient is absent in 
the case of a super-lens geometry \cite{Kastel04}.
To illustrate the theory, we apply it to the case of the
two atoms being near a realistic dielectric microsphere.  
The formalism used is based on the quantization
of the macroscopic electromagnetic field
and allows to take into account material dispersion
and absorption in a quantum-mechanically consistent manner.

The paper is organized as follows. In Sec.~\ref{sec2}
the basic equations for describing the interaction
of $N$ multilevel atoms with the electromagnetic field 
in the presence of dispersing and absorbing macroscopic bodies 
are given. In Sec.~\ref{sec3} the theory is applied to 
the problem of formation of an entangled state between two
$\Lambda$-type three-level atoms.
Section \ref{sec5} presents the results obtained for the
case when the two atoms are at diametrically opposite
positions outside a microsphere. Finally, a summary
and some concluding remarks are given in Sec.~\ref{sec6}. 


\section{Master equation}
\label{sec2}

Consider $N$ multilevel atoms
at given positions
$\mathbf{r}_A$ 
that interact with the electromagnetic field 
in the presence 
of some macroscopic, linear bodies, which
are allowed to be both dispersing and absorbing.
In electric dipole approximation, the overall system
can be described by the multipolar-coupling Hamiltonian \cite{ksw},
\begin{align}
\label{e2}
   &\hat{H} = \int {\rm d}^3{\bf r}
   \int_0^\infty {\rm d}\omega \,\hbar\omega
   \,\hat{\bf f}^\dagger({\bf r},\omega){}\hat{\bf f}({\bf r},\omega)
\nonumber\\[1ex]
    &+\!\sum_{A} \sum_m \hbar \omega_{Am} \hat{R}_{Amm}
    \!-\! \sum_{A}\! \int_0^\infty\!\! {\rm d}\omega \bigl[
    \hat{\bf d}_A
    \underline{\hat{\bf E}}({\bf r}_A,\omega)
    + {\rm H.c.}\bigr].
\end{align}
Here, the bosonic fields $\hat{\mathbf{f}}(\mathbf{r},\omega)$ 
and $\hat{\mathbf{f}}^\dagger(\mathbf{r},\omega)$,
\begin{equation}
\label{e2-1}
\bigl[\hat{f}_k(\mathbf{r},\omega),
\hat{f}_{k'}^\dagger(\mathbf{r}',\omega')\bigr]
= 
\delta_{kk'}
\delta(\omega-\omega')\delta(\mathbf{r}-\mathbf{r}'),
\end{equation}
are the canonically conjugated variables of the system, which
consists of the electromagnetic field and the bodies (including the
dissipative system responsible for absorption),
the $\hat{R}_{Amn}$ are the atomic (flip) operators
\begin{equation}
\label{e2.2}
      \hat{R}_{Amn} = |m\rangle_A{}_A\langle n|,
\end{equation}
with $|m\rangle_A$ being the $m$th energy eigenstate of the
$A$th atom (of energy $\hbar\omega_{Am}$), and 
\begin{equation}
\label{e2.1}
        \hat{\bf d}_A = \sum_{m,n} {\bf d}_{Amn} 
        \hat{R}_{Amn}
\end{equation}
are the electric dipole operators of the atoms (${\bf d}_{Amn}$
$\!=$ $\!_A\langle m|\hat{\mathbf{d}}_A|n\rangle_A$).
Further, the body-assisted electric field in the $\omega$ domain,
$\underline{\hat{\bf E}}({\bf r},\omega)$, expressed
in terms of the fundamental variables
$\hat{\mathbf{f}}(\mathbf{r},\omega)$ reads
\begin{equation}
\label{e3}
     \underline{\hat{\bf E}}({\bf r},\omega)
     = \int {\rm d}^3{\bf r}'\,
     \tilde{\bm{G}}({\bf r},{\bf r}',\omega)
     \hat{\bf f}({\bf r}',\omega),
\end{equation}
where
\begin{equation}
\label{e3.1}
    \tilde{\bm{G}}({\bf r},{\bf r}',\omega)
    = i \sqrt{\frac{\hbar}{\pi\varepsilon_0}}
    \frac{\omega^2}{c^2}
    \sqrt{\mathrm{Im}\,\varepsilon({\bf r}',\omega)}
    \,\bm{G}({\bf r},{\bf r}',\omega),
\end{equation}
with
$\bm{G}(\rb,\rb',\omega)$
being
the classical Green tensor which satisfies the equation
\begin{equation}
\label{e}
    \bm{\nabla}\times
    \bm{\nabla}\times\bm{G}(\rb,\rb',\omega)-\frac{\omega^2}   
    {c^2}\varepsilon(\rb,\omega)\bm{G}(\rb,\rb',\omega)
    =\bm{\delta}(\rb-\rb')
\end{equation}
together with the boundary conditions at infinity
[$\bm{\delta}(\rb)$, dyadic $\delta$ function]. Throughout
the paper we restrict our attention to dielectric bodies, which are
described by a spatially varying complex permittivity
$\varepsilon(\mathbf{r},\omega)$ $\!=$ $\!\mathrm{Re}\,
\varepsilon(\mathbf{r},\omega)$ $\!+$ $\!i\mathrm{Im}\,
\varepsilon(\mathbf{r},\omega)$.

Next we assume that the macroscopic bodies, say, microspheres or 
photonic crystals, act like resonator-like equipments such that
the excitation spectrum of the body-assisted electromagnetic-field
shows a resonance structure, with the lines being well separated 
from each other. With regard to the atom--field coupling, we
assume that a few atomic transitions can be strongly coupled to
field resonances tuned to them, while all other transitions are  
weakly coupled to the field.
Following Ref.~\cite{Ho02}, we decompose the body-assisted
electromagnetic field into the part
(denoted by $\ints {\rm d\omega}\ldots$)
that can be strongly coupled to atomic transitions and the rest
(denoted by $\intss {\rm d\omega}\ldots$), which
only gives rise to a weak atom--field coupling.   
The Heisenberg equation of motion for 
an arbitrary operator $\hat{O}$
that belongs to the system consisting of the atoms and
the part of the body-assisted electromagnetic field
that strongly interacts with the atoms
can then be written in the form of
\begin{eqnarray}
\label{e4}
\lefteqn{
       \dot{\!\hat{O}} =
       -\frac{i}{\hbar}
       \bigl[\hat{O},\hat{H}\bigr] =
       -\frac{i}{\hbar}
       \bigl[\hat{O},\hat{H}_{\rm S}\bigr]
}
\nonumber\\[1ex]&&\hspace{1ex}
       +\,\frac{i}{\hbar} \sum_A \Intss {\rm d}\omega\,
       \left\{\bigl[\hat{O},\hat{\bf d}_A\bigr]
       \underline{\hat{\bf E}}({\bf  r}_A,\omega)
       \right.
\nonumber\\[1ex]&&\hspace{20ex}
       \left.
       +\,\underline{\hat{\bf E}}^\dagger({\bf  r}_A,\omega)
       \bigl[\hat{O},\hat{\bf d}_A\bigr]\right\},
\end{eqnarray}
where
\begin{align}
\label{e5}
   &\hat{H}_{\rm S} = \int {\rm d}^3{\bf r}
   \Ints {\rm d}\omega \,\hbar\omega
   \,\hat{\bf f}^\dagger({\bf r},\omega){}\hat{\bf f}({\bf r},\omega)
\nonumber\\[1ex]
    &+\! \sum_A 
      \sum_m \hbar\omega_{Am} \hat{R}_{Amm}
    \!-\! \sum_{A} \!\Ints \!{\rm d}\omega \bigl[
    \hat{\bf d}_A
    \underline{\hat{\bf E}}({\bf r}_A,\omega)
    + {\rm H.c.}\bigr].    
\end{align}

To handle the weak atom--field interaction,
i.e., the integral $\intss {\rm d\omega}\ldots$ in
Eq.~(\ref{e4}), we first formally solve the  
Heisenberg equation of motion 
\begin{align}
\label{e6}
       \dot{\hat{\!\bf f}}({\bf r},\omega)
       & = -\frac{i}{\hbar}
       \bigl[\hat{{\bf f}}({\bf r},\omega),\hat{H}\bigr]
\nonumber\\[1ex]       
       & = -i\omega \hat{{\bf f}}({\bf r},\omega)
       +\frac{i}{\hbar} \sum_{A} \hat{{\bf d}}_{A}
       \tilde{\bm{G}}^\ast({\bf r}_{A},{\bf r},\omega),
\end{align}
which yields
\begin{eqnarray}
\label{A1}
\lefteqn{
       \hat{\bf f}({\bf r},\omega,t)
       = \hat{\bf f}_{\rm free}({\bf r},\omega,t)
}
\nonumber\\[1ex]&&\hspace{0ex}
       +\,
\frac{i}{\hbar} 
        \sum_{A} \int_0^t {\rm d} t'\,
        \hat{\bf d}_{A}(t')
        \,\tilde{\bm{G}}^\ast({\bf r}_{A},{\bf r},\omega)
        e^{-i\omega(t-t')},
\end{eqnarray}
where $\hat{\bf f}_{\rm free}({\bf r},\omega,t)$ evolves freely,
\begin{equation}
\label{A1-1}
      \hat{\bf f}_{\rm free}({\bf r},\omega,t)
      = \hat{\bf f}_{\rm free}({\bf r},\omega,0) 
      e^{-i\omega t}.
\end{equation}
Inserting Eq.~(\ref{A1}) into Eq.~(\ref{e3}), we derive
\begin{eqnarray}
\label{A1-2}
\lefteqn{
       \underline{\hat{\bf E}}({\bf  r},\omega,t) =
       \underline{\hat{\bf E}}_{\rm free}({\bf  r},\omega,t)
}
\nonumber\\[1ex]&&
        \!\!\!+\,
        \frac{i}{\pi\varepsilon_0}\frac{\omega^2}{c^2}
        \sum_A \!\int_0^t \!{\rm d} t'\,
        e^{-i\omega(t-t')}
        {\rm Im}\, \bm{G}({\bf r},{\bf r}_A,\omega)
        \,\hat{\bf d}_A(t'), \qquad
\end{eqnarray}
where $\underline{\hat{\bf E}}_{\rm free}({\bf  r},\omega,t)$ is
defined according to Eq.~(\ref{e3}) with
$\hat{\bf f}_{\rm free}({\bf r},\omega,t)$ in place of
$\hat{\bf f}({\bf r},\omega,t)$.
Introducing slowly varying atomic operators
\begin{gather}
\label{A1-3}
   \hat{\tilde{R}}_{Amn}(t)
   = \hat{R}_{Amn}(t) e^{-i\tilde{\omega}_{Amn}t},
\\[1ex]
\label{e8.0}
        \tilde{\omega}_{Amn} = \tilde{\omega}_{Am}-
        \tilde{\omega}_{An},
\end{gather}
where the $\hbar\tilde{\omega}_{An}$ are the atomic 
energy levels including the anticipated media-induced shifts,
we may write the electric dipole operator,
Eq.~(\ref{e2.1}), as
\begin{eqnarray}
\label{e7}
&\displaystyle
        \hat{\bf d}_A(t) = \sum_{m,n} {\bf d}_{Amn} 
        \hat{\tilde{R}}_{Amn}(t) e^{i\tilde{\omega}_{Amn} t}.
\end{eqnarray}

We now insert Eq.~(\ref{A1-2}) together with Eq.~(\ref{e7})
in the integral $\intss {\rm d\omega}\ldots$
in Eq.~(\ref{e4}), apply the Markov approximation to the
slowly varying atomic variables in the time integral,
and take the expectation value. Assuming that the
free field is initially in the vacuum state, we derive 
(cf. App.~A of Ref.~\cite{Ho02})
\begin{eqnarray}
\label{e8}
\lefteqn{
       \bigl\langle\, \dot{\hat{O}} \bigr\rangle
       = -\frac{i}{\hbar}
       \bigl\langle \bigl[\hat{O},\,\hat{\!\tilde{H}}_{\rm S}
       \bigr] \bigr\rangle
}
\nonumber\\[1ex]&&\hspace{0ex}
       +\, i\sum_{A,A'}
       \sumprime
       \sum_{m,n}
       \Bigl( \delta^{mn}_{AA^{'}} 
       \bigl\langle\bigl[\hat{O},\hat{R}_{Amn}\bigr]
       \hat{R}_{A'nm}\bigr\rangle
\nonumber\\[1ex]&&\hspace{5ex}
       +\, \delta^{nm}_{A A^{'}} 
       \bigl\langle\hat{R}_{A'nm} \bigl[\hat{O},\hat{R}_{Amn}\bigr]
       \bigr\rangle\Bigr)
\nonumber\\[1ex]&&\hspace{0ex}
       -\, \frac{1}{2} \sum_{A,A'} \sum_{m,n} \sumprime
       \Bigl( \Gamma^{mn}_{AA'} 
       \bigl\langle\bigl[\hat{O},\hat{R}_{Amn}\bigr]
       \hat{R}_{A'nm}\bigr\rangle
\nonumber\\[1ex]&&\hspace{5ex}
       -\, \Gamma^{nm}_{A A^{'}} 
       \bigl\langle\hat{R}_{A'nm} \bigl[\hat{O},\hat{R}_{Amn}\bigr]
       \bigr\rangle\Bigr),
\end{eqnarray}
where the primed sum $\sum_{A,A'}'$ 
indicates that \mbox{$A$ $\!\neq$ $A'$} and
the primed sum $\sum_{m,n}'$ indicates that
transitions that can strongly interact with
the body-assisted electromagnetic field are excluded.
In Eq.~(\ref{e8}), $\hat{\tilde{H}}_{\rm S}$ is defined 
according to Eq.~(\ref{e5}), with $\omega_{Am}$ being replaced by  
\begin{equation}
\label{e11}
     \tilde{\omega}_{Am} = 
      \omega_{Am} - \delta^m_{AA},
\end{equation}
where
\begin{equation}
\label{e12}
      \delta^m_{AA} = \sum_n \delta^{mn}_{AA},
\end{equation}
with $\delta^{mn}_{AA}$ being obtained from 
\begin{align}
\label{e10}
&\delta^{mn}_{AA^{'}} =
\frac{1}{\hbar\pi\varepsilon_0c^2}\,
{\cal P}\!\int_0^\infty {\rm d}\omega \omega^2
\nonumber\\[1ex]
&\hspace{8ex}\times\,
     \frac{
     {\bf d}_{Amn} \,{\rm Im}\,\bm{G}({\bf r}_A,{\bf r}_{A'},
     \omega)\,
     {\bf d}^\ast_{A'mn}
     }
      {\omega-\tilde{\omega}_{A'mn}}
\end{align}
($\mathcal{P}$, principal part) for $A$ $\!=$ $\!A'$. For
$A$ $\!\neq$ $\!A'$, the parameters $\delta^{mn}_{AA^{'}}$ are
the dipole--dipole coupling strengths between different atoms $A$
and $A'$. Further, the decay rates $\Gamma^{mn}_{AA^{'}}$ 
are defined according to
\begin{align}
\label{e9}
\Gamma^{mn}_{AA^{'}} = &\
\frac{2\tilde{\omega}_{A'mn}^2}{\hbar\varepsilon_0c^2}
     \,\Theta(\tilde{\omega}_{A'mn})
\nonumber\\[1ex]
&\times\,
     {\bf d}_{Amn} \,{\rm Im}\,\bm{G}({\bf r}_A,{\bf r}_{A'},
     \tilde{\omega}_{A'mn})\,
     {\bf d}^\ast_{A'mn}
\end{align}
[$\Theta(x)$, unit step function]. 

Using the relationship
\begin{align}
\label{e13}
      \bigl\langle \hat{O}(t) \bigr\rangle
      &=
      {\rm Tr} \bigl[\hat{\rho}(0) \hat{O}(t)\bigr]
\nonumber\\[1ex]
      &= {\rm Tr} \bigl[\hat{\rho}(t) \hat{O}(0)\bigr]
      = {\rm Tr} \bigl[\hat{\varrho}(t) \hat{O}(0)\bigr],
\end{align}
where $\hat{\rho}$ is the density operator of the overall system, 
and $\hat{\varrho}$ is the (reduced) density operator of the
system under consideration, and making use of the cyclic properties 
of the trace, from Eq.~(\ref{e8}) we derive the following
equation of motion for the system density operator in
the Schr\"odinger picture:
\begin{align}
\label{e14}
&
       \,\dot{\!\hat{\varrho}} =
       -\frac{i}{\hbar}
       \bigl[\,\hat{\!\tilde{H}}_{\rm S},
       \hat{\varrho}\bigr]
       + 
       \biggl[i
       \sum_{A,A'}\sumprime \sum_{m,n}
       \delta^{mn}_{AA^{'}} 
       (\hat{R}_{Amn} \hat{R}_{A'nm} \hat{\varrho}
\nonumber\\
&\hspace{20ex}
       - \hat{R}_{A'nm} \hat{\varrho} \hat{R}_{Amn})
       + {\rm H.c.} \biggr]
\nonumber\\[1ex]
&\hspace{5ex}
       - \frac{1}{2}  \sum_{A,A'} \sum_{m,n} \sumprime
       \Bigl[ \Gamma^{mn}_{AA^{'}} 
       (\hat{R}_{Amn} \hat{R}_{A'nm} \hat{\varrho}
\nonumber\\
&\hspace{10ex}       
       - \hat{R}_{A'nm} \hat{\varrho} \hat{R}_{Amn})
       + {\rm H.c.} \Bigr].
\end{align}
Equation (\ref{e14}) is a generalization of the two-level-atom 
result in Ref.~\cite{Ho02} to the case of multilevel atoms.
In particular, if the conditions
\begin{align}
\label{e16}
       \delta^{mn}_{AA^{'}} &= \delta^{mn}_{A^{'}A},
\\[1ex]
\label{e15}
       \Gamma^{mn}_{AA^{'}} &= \Gamma^{mn}_{A^{'}A}
\end{align}
are fulfilled, which is the case when, for example, the atoms are 
identical and located in free space or at equivalent positions 
with respect to the macroscopic bodies, then the master 
equation (\ref{e14}) takes the somewhat simpler form of 
\begin{align}
\label{e17}
       \,\dot{\!\hat{\varrho}} =
       & -\frac{i}{\hbar} 
        \left[\,\hat{\!\tilde{H}}_{\rm S}
         +
       \hat{H}_{\rm D}
       ,
       \hat{\varrho}\right]
\nonumber\\[1ex]
       &- \frac{1}{2} \sum_{A,A'} \sum_{m,n} \sumprime
       \Gamma^{mn}_{AA^{'\ast}} 
       (\hat{R}_{Amn} \hat{R}_{A'nm} \hat{\varrho}
\nonumber\\[1ex]
       &- 2\hat{R}_{A'nm} \hat{\varrho} \hat{R}_{Amn}
       + \hat{\varrho} \hat{R}_{Amn} \hat{R}_{A'nm} ),
\end{align}
where
\begin{eqnarray}
\label{e18}
    \hat{H}_{\rm D}
    = 
    -  \sum_{A,A'}\sumprime \sum_{m>n}
       \hbar \Delta^{mn}_{AA'} 
       \hat{R}_{Amn} \hat{R}_{A'nm}
\end{eqnarray}
describes the dipole-dipole interaction between the atoms, with
$\Delta^{mn}_{AA'}$ being the dipole-dipole coupling strengths,
\begin{equation}
\label{e19}
     \Delta^{mn}_{AA'} = \delta^{mn}_{AA^{'}}
          + \delta^{nm}_{A'A}.
\end{equation}
According to Eq.~(\ref{e17}), the (undamped) system is
governed by an effective Hamiltonian equal to 
$\hat{\tilde{H}}_{\rm S}$ $\!+$ $\!\hat{H}_{\rm D}$.
Note that this is not true in general, but only under 
the conditions (\ref{e16}) and (\ref{e15}).

To construct the (formal) solution to the master equation 
(\ref{e17}), we first rewrite it in the form of
\begin{equation}
\label{e20}
     \dot{\!\hat{\varrho}} = \hat{L} \hat{\varrho}
      + \hat{S} \hat{\varrho},
\end{equation}
where $\hat{L}$ and $\hat{S}$ are superoperators which act 
on $\hat{\varrho}$ according to the rules
\begin{gather}
\label{e21}
      \hat{L} \hat{\varrho} \equiv - \frac{i}{\hbar} 
      (
      \hat{\mathcal{H}}
      \hat{\varrho} 
      - \hat{\varrho}
      \hat{\mathcal{H}}^\dagger
      ),
\\[1ex]
\label{e22}
    \hat{S} \hat{\varrho} \equiv 
       \sum_{A,A'} \sum_{m,n} \sumprime
       \Gamma^{mn}_{AA^{'}} 
       \hat{R}_{A'nm} \hat{\varrho} \hat{R}_{Amn},
\end{gather}
and the non-Hermitian ``Hamiltonian'' $\hat{\mathcal{H}}$ reads
\begin{eqnarray}
\label{e23}
       \hat{\mathcal{H}}
       = \hat{\!\tilde{H}}_{\rm S}
       + \hat{H}_{\rm D}
       - \frac{i\hbar}{2} \sum_{A,A'} \sum_{m>n} \sumprime
       \Gamma^{mn}_{AA^{'}} 
       \hat{R}_{Amn} \hat{R}_{A'nm}.
\end{eqnarray}
{F}rom Eqs.~(\ref{e20})--(\ref{e22}) it then follows that
\begin{eqnarray}
\label{e24}
     \hat{\varrho}(t) = e^{\hat{L}(t-t_0)}\hat{\varrho}(t_0)
     + \int_{t_0}^t {\rm d}t_1 e^{\hat{L}(t-t_1)}\hat{S}
       \hat{\varrho}(t_1).
\end{eqnarray}
By iteration, from Eq. (\ref{e24}) one readily finds
\begin{equation}
\label{e25}
     \hat{\varrho}(t) = \sum_{n=0}^\infty 
     \hat{\varrho}^{(n)}(t),
\end{equation}
where
\begin{align}
\label{e25.1}
     &\hat{\varrho}^{(0)}(t) = 
      e^{\hat{L}(t-t_0)}\hat{\varrho}(t_0),
\\[1ex]
\label{e25.2}
     &\hat{\varrho}^{(n)}(t) = 
       \int_{t_0}^t {\rm d}t_n
       \int_{t_0}^{t_n} {\rm d}t_{n-1}\ldots
       \int_{t_0}^{t_2} {\rm d}t_1 
       e^{\hat{L}(t-t_n)}
\nonumber\\[1ex]
      &\hspace{1ex}\times
       \hat{S}e^{\hat{L}(t_n-t_{n-1})}
       \ldots
       \hat{S}e^{\hat{L}(t_1-t_0)}
       \hat{\varrho}(t_0), \ n=1,2,3\ldots.
\end{align}
Although Eq. (\ref{e25}) is not a perturbative expansion,
it can be helpful, as we shall see below, in finding the 
explicit solutions to the master equation.


\section{Two three-level atoms of $\bm{\Lambda}$ type}
\label{sec3}

\subsection{Solution to the master equation}
\label{sec3A}

\begin{figure}[htb]

\noindent
\begin{center}
\includegraphics[width=.67\linewidth]{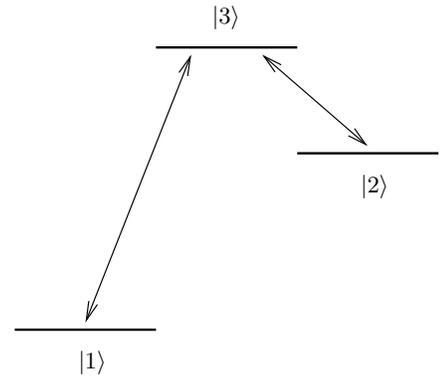}
\end{center}
\caption{Three-level atom of $\Lambda$ type.}
\label{fig1}
\end{figure}

Let us specify the atomic system and consider two identical
three-level atoms $A$ and $B$ of $\Lambda$ type as 
sketched in Fig.~\ref{fig1}.
We assume that the dipole-allowed transition
$|1\rangle$ $\!\leftrightarrow$ $\!|3\rangle$ is
tuned to a well pronounced body-induced electromagnetic 
field resonance, thereby giving rise to a strong 
dipole-allowed atom--field coupling. Further, 
the dipole-allowed transition 
\mbox{$|2\rangle$ $\!\leftrightarrow$ $\!|3\rangle$} 
is assumed to be weakly coupled to the body-assisted electromagnetic
field, and the transition between the states $|1\rangle$ and
$|2\rangle$ is dipole-forbidden.
Restricting our attention to 
two atoms at equivalent positions with respect
to the macroscopic bodies, so that 
corresponding transition frequencies are equally
shifted and the relations
\begin{align}
\label{e35}
\Delta_{AB}^{31} =& \;\Delta_{BA}^{31},
&\hspace{-10ex}
\Delta_{AB}^{32} =& \;\Delta_{BA}^{32},
\\[1ex]
\label{e36}
\Gamma_{AA}^{32} =& \;\Gamma_{BB}^{32}, 
&\hspace{-10ex}
\Gamma_{AB}^{32} =& \;\Gamma_{BA}^{32} 
\end{align}
hold [cf. Eqs.~(\ref{e16}) and (\ref{e15})],
we may apply the master equation
in the form of Eq.~(\ref{e17}) and its solution in the form 
of Eqs.~(\ref{e25})--(\ref{e25.2}), with Eqs.~(\ref{e22})
and (\ref{e23}) being explicitly given by
\begin{equation}
\label{e37}
    \hat{S} \hat{\varrho} \equiv 
       \sum_{A',A''=A,B} 
       \Gamma^{32}_{A'A''} 
       \hat{R}_{A''23} \hat{\varrho} \hat{R}_{A'32}
\end{equation}
and 
\begin{equation}
\label{e37.1}
       \hat{\mathcal{H}}
       = \hat{\!\tilde{H}}_{\rm S}
       + \hat{H}_{\rm D}
       - \frac{i\hbar}{2} \sum_{A',A''}
       \Gamma^{32}_{A'A''} 
       \hat{R}_{A'32} \hat{R}_{A''23},
\end{equation}
where
\begin{align}
\label{e37.2}
   \hat{\tilde{H}}_{\rm S} &= \int {\rm d}^3{\bf r}
   \Ints {\rm d}\omega \,\hbar\omega
   \,\hat{\bf f}^\dagger({\bf r},\omega)
   {}\hat{\bf f}({\bf r},\omega)
\nonumber\\
    &+ \sum_{A'=A,B} 
      \sum_{m=1}^3 \hbar
      \tilde{\omega}_{A'm} \hat{R}_{A'mm}
\nonumber\\
    &- \sum_{A'=A,B} \!\Ints \!{\rm d}\omega \bigl[
    {\bf d}_{A'31} \hat{R}_{A'31}
    \underline{\hat{\bf E}}({\bf r}_{A'},\omega)
    + {\rm H.c.}\bigr],
\\[1ex]
\label{e37.3}
    \hat{H}_{\rm D}
    &= 
    -  (\hbar \Delta^{31}_{AB} 
       \hat{R}_{A31} \hat{R}_{B13}
       + \Delta^{32}_{AB} 
       \hat{R}_{A32} \hat{R}_{B23}) + {\rm H.c.},
\end{align}
with the rotating-wave approximation in Eq.~(\ref{e37.2}).

To specify the initial condition at time $t_0$, let us assume
that the two atoms are initially in the ground state $|1,1\rangle$ 
($|i,j\rangle$ $\!\equiv$ $\!|i\rangle_A\otimes|j\rangle_B$,
$i,j$ $\!=$ $\!1,2,3$) and the rest of the system
is prepared in a state
\begin{equation}
\label{e38}
      |F\rangle = 
      \Ints \mathrm{d}\omega
      \int \mathrm{d}^3\mathbf{r}\,
      \mathbf{C}(\mathbf{r},\omega,t_0)
      \hat{\mathbf{f}}^\dagger
      (\mathbf{r},\omega)|\{0\}\rangle,
\end{equation}
where $\mathbf{C}(\mathbf{r},\omega,t_0)$ as a function of 
$\omega$ is non-zero in a small interval around 
$\omega$ $\!\simeq$ $\!\tilde{\omega}_{A31}$
$\!=$ $\!\tilde{\omega}_{B31}$, and $|\{0\}\rangle$ 
is vacuum state with respect to this frequency interval.  
The initial density operator can then be given in the form of
\mbox{($t_0$ $\!=$ $\!0$)}
\begin{equation}
\label{e39}
   \hat{\varrho}(0) = |\Psi(0)\rangle\langle\Psi(0)|,
   \qquad
   |\Psi(0)\rangle = |1,1\rangle \otimes|F\rangle.
\end{equation}
In order to determine the density operator at time $t$, we
begin by calculating the first term of the series 
(\ref{e25}), viz.
\begin{equation}
\label{e41}
     \hat{\varrho}^{(0)}(t)
      = e^{\hat{L}t}\hat{\varrho}(0)
      = |\Psi(t)\rangle\langle\Psi(t)|,
\end{equation}
where the (damped) state vector
\begin{equation}
\label{e42}
|\Psi(t)\rangle = e^{-i\hat{\mathcal{H}}t/\hbar}|\Psi(0)\rangle
\end{equation}
obviously obeys the equation
\begin{equation}
\label{e43}
    i\hbar\,\frac{\mathrm{d}|\Psi(t)\rangle}{\mathrm{d}t}
    = \hat{\mathcal{H}}|\Psi(t)\rangle.
\end{equation}
Recalling the initial condition (\ref{e39}) and the form
of $\hat{\mathcal{H}}$, Eqs.~(\ref{e37.1})--(\ref{e37.3}), we may 
expand $|\Psi(t)\rangle$ as
\begin{align}
\label{e44}
    &|\Psi(t)\rangle =
    C_{31}(t)e^{-i(\tilde{\omega}_{A1}+\tilde{\omega}_{B3})t}
    |3,1\rangle\otimes|\{0\}\rangle
\nonumber\\[1ex]
&\hspace{8ex}
    +C_{13}(t)e^{-i(\tilde{\omega}_{A3}+\tilde{\omega}_{B1})t}
    |1,3\rangle\otimes|\{0\}\rangle
\nonumber\\[1ex]
&\hspace{8ex}     
    +\Ints \mathrm{d}\omega
    \int \mathrm{d}^3\mathbf{r}\,
    e^{-i(\tilde{\omega}_{A1}+\tilde{\omega}_{B1}+\omega) t}
\nonumber\\[1ex]
    &\hspace{12ex}\times\,
    \mathbf{C}(\mathbf{r},\omega,t)
    \hat{\mathbf{f}}^\dagger(\mathbf{r},\omega)|\{0\}\rangle
    \otimes|1,1\rangle.
\end{align}
We now substitute Eq.~(\ref{e44}) into Eq.~(\ref{e43}) and make 
explicitly use of Eqs.~(\ref{e37.1})--(\ref{e37.3}).
Straightforward calculation yields the following system of 
differential equations for the expansion coefficients:
\begin{widetext}
\begin{align}
\label{e45}
    &\dot{C}_{31} = 
    -{\textstyle\frac{1}{2}}\Gamma^{32}_{AA} C_{31}
    + i\Delta_{AB}^{31} C_{13}
    + \frac{i}{\hbar}
    \Ints \mathrm{d}\omega \int \mathrm{d}^3\mathbf{r}\, 
    \db_{A31}\tilde{\bm{G}}(\rb_A,\rb,\omega){\bf C}
    (\rb,\omega)e^{-i(\omega-\tilde\omega_{A31})t},
\\[1ex]
\label{e46}
    &\dot{C}_{13} = 
      -{\textstyle\frac{1}{2}}\Gamma^{32}_{BB} C_{13}
    + i\Delta_{AB}^{31} C_{31}
    + \frac{i}{\hbar}
     \Ints \mathrm{d}\omega \int \mathrm{d}^3\mathbf{r}\, 
    \db_{B31}\tilde{\bm{G}}(\rb_B,\rb,\omega){\bf C}
    (\rb,\omega)e^{-i(\omega-\tilde\omega_{B31})t},
\\[1ex]
\label{e47}
     &\dot{{\bf C}}(\rb,\omega) =
    \frac{i}{\hbar}
    e^{i(\omega-\tilde{\omega}_{A31})t}
    [\db^\ast_{A31}
    \tilde{\bm{G}}^*(\rb_A,\rb,\omega)C_{31}
    + \db^\ast_{B31}
    \tilde{\bm{G}}^*(\rb_B,\rb,\omega)C_{13}].
\end{align}
\end{widetext}
Recall that $\tilde{\omega}_{A31}$ $\!=$ $\!\tilde{\omega}_{B31}$.
Inserting the formal solution to Eq.~(\ref{e47}) in Eqs.~(\ref{e45}) 
and (\ref{e46}), we derive, on making use of the properties of the 
Green tensor, the integro-differential equations
\begin{align}
\label{e48}
      \dot{C}_{31} = &-{\textstyle\frac{1}{2}}\Gamma^{32}_{AA} C_{31}
    + i\Delta_{AB}^{31} C_{13}
\nonumber\\[1ex]
     &+ \int_0^t \mathrm{d}t'\,
     [K_{AA}(t-t')C_{31}(t')
\nonumber\\[1ex]
      &+ K_{AB}(t-t')C_{13}(t')]
     + F_{31}(t),
\end{align}
\begin{align}
\label{e48.1}
      \dot{C}_{13} =& -{\textstyle\frac{1}{2}}\Gamma^{32}_{BB} C_{13}
    + i\Delta_{AB}^{31} C_{31}
\nonumber\\[1ex]
     +& \int_0^t \mathrm{d}t'\,
     [K_{BB}(t-t')C_{13}(t')
\nonumber\\[1ex]
      +& K_{BA}(t-t')C_{31}(t')]
     + F_{13}(t),
\end{align}
where the kernel function $K_{A'A''}(t)$ is defined by
\begin{align}
\label{e49}
     &K_{A'A''}(t) = - \frac{1}{\hbar\pi\varepsilon_0}
     \Ints \mathrm{d}\omega
     \,\frac{\omega^2}{c^2}\,
      e^{-i(\omega-\tilde{\omega}_{A31})t}
\nonumber\\[1ex]
     &\hspace{15ex}\times
      \mathbf{d}_{A'31}\,\mathrm{Im}\,
     \bm{G}(\mathbf{r}_{A'},\mathbf{r}_{A''},\omega)
\,\mathbf{d}^\ast_{A''31}
\end{align}
[$A'(A'')$ $\!=$ $\!A,B$], and the free-field driving terms 
$F_{31}$ and $F_{13}$ read
\begin{align}
\label{e50}
     &F_{31}(t) =
     \frac{i}{\hbar}
     \Ints \mathrm{d}\omega 
      \int \mathrm{d}^3\mathbf{r}\,
      \db_{A31}\tilde{\bm{G}}(\rb_A,\rb,\omega)
\nonumber\\[1ex]
&\hspace{10ex}\times
      {\bf C}(\rb,\omega,0)
      e^{-i(\omega-\tilde\omega_{A31})t},
\end{align}
\begin{align}
\label{e50.1}
     &F_{13}(t) =
     \frac{i}{\hbar}
     \Ints \mathrm{d}\omega 
      \int \mathrm{d}^3\mathbf{r}\,
      \db_{B31}\tilde{\bm{G}}(\rb_B,\rb,\omega)
\nonumber\\[1ex]
&\hspace{10ex}\times
      {\bf C}(\rb,\omega,0)
      e^{-i(\omega-\tilde\omega_{B31})t}.
\end{align}
Note that for identical atoms at equivalent positions
with respect to the macroscopic bodies
\begin{equation}
\label{e50-0}
       K_{AA}(t) = K_{BB}(t),
       \quad
       K_{AB}(t) = K_{BA}(t).
\end{equation}
Instead of considering the probability amplitudes 
$C_{31}$ and $C_{13}$, it is advantageous to introduce the
probability amplitudes
\begin{equation}
\label{e50-1}
      C_\pm^{13} = 2^{-\frac{1}{2}}\left(C_{31} \pm 
      C_{13}\right),
\end{equation}
which are the expansion coefficients of $|\Psi\rangle$
with respect to the atomic basis
\begin{equation}
\label{e50-2}
   |\pm_{13}\rangle = 
   2^{-\frac{1}{2}}\left(|3,1\rangle \pm |1,3\rangle\right),
\end{equation}
so that Eq.~(\ref{e44}) takes the form of
\begin{align}
\label{e55}
    &|\Psi(t)\rangle =
    C_+^{13}(t)e^{-i(\tilde{\omega}_{A1}+\tilde{\omega}_{B3})t}
    |+_{13}\rangle\otimes|\{0\}\rangle
\nonumber\\[1ex]
&\hspace{8ex}
    +C_-^{13}(t)e^{-i(\tilde{\omega}_{A1}+\tilde{\omega}_{B3})t}
    |-_{13}\rangle\otimes|\{0\}\rangle
\nonumber\\[1ex]
&\hspace{8ex}
    +\Ints \!\!\mathrm{d}\omega
    \!\int \!\mathrm{d}^3\mathbf{r}\,
    e^{-i(\tilde{\omega}_{A1}+\tilde{\omega}_{B1}+\omega) t}
\nonumber\\[1ex]
    &\hspace{12ex}\times
    \mathbf{C}(\mathbf{r},\omega,t)
    \hat{\mathbf{f}}^\dagger(\mathbf{r},\omega)|\{0\}\rangle
    \otimes|1,1\rangle.
\end{align}
{F}rom Eqs.~(\ref{e48})--(\ref{e50-1}) it is not difficult to see 
that the differential equations for $C_\pm^{13}$ decouple
\begin{align}
\label{e50-3}
   \dot{C}_\pm^{13}
   =& \left(\pm i\Delta_{AB}^{31}
   -{\textstyle\frac{1}{2}}\Gamma^{32}_{AA}
   \right) C_\pm^{13} 
\nonumber\\[1ex]
   &+ \int_0^t \mathrm{d}t'\,
   K_\pm(t-t')C_\pm^{13}(t') + F_\pm(t),
\end{align}
where
\begin{eqnarray}
\label{e50-4}
&\displaystyle
      K_\pm(t) = K_{AA}(t) \pm K_{AB}(t),
\\[1ex]
\label{e50-5}
&\displaystyle
      F_\pm(t) = 2^{-1/2}[F_{31}(t)\pm F_{13}(t)].
\end{eqnarray}

The field resonance strongly coupled to the atomic transition
\mbox{$|1\rangle$ $\!\leftrightarrow$ $\!|3\rangle$}
can be typically modeled by a Lorentzian,
with $\omega_\mathrm{C}$ $\!\approx$ $\!\tilde{\omega}_{A31}$
and $\Delta\omega_\mathrm{C}$ being the central frequency 
and the half width at half maximum, respectively. In this case, 
Eq.~(\ref{e49}) can be approximated by 
\begin{align}
\label{e51}
    K_{A'A''}(t) = &-\Gamma_{A'A''}^{31}\,
    e^{-i(\omega_\mathrm{C}-\tilde{\omega}_{A31})t}
\nonumber\\[1ex]
&\times\,
    \frac{1}{2\pi}\int \mathrm{d}\omega\,
    \frac{\Delta\omega_\mathrm{C}^2
    e^{-i(\omega-\omega_\mathrm{C})t}}
    {(\omega-\omega_\mathrm{C})^2
    +\Delta\omega_\mathrm{C}^2}\,,
\end{align}
where $\Gamma_{A'A''}^{31}$ is defined according to 
Eq.~(\ref{e9}), but with $\tilde{\omega}_{A31}$ being 
replaced by $\omega_\mathrm{C}$,
\begin{equation}
\label{e52}
     \Gamma^{31}_{A'A''} = 
     \frac{2\omega_\mathrm{C}^2}
     {\hbar\varepsilon_0c^2}\,
     {\bf d}_{A'31} \,{\rm Im}\,
     \bm{G}({\bf r}_{A'},{\bf r}_{A''},
     \omega_\mathrm{C})\,
     {\bf d}^\ast_{A''31}.
\end{equation}
{F}rom Eq.~(\ref{e51}) it then follows that ($t$ $\!\ge$ $\!0$)
\begin{equation}
\label{e53}
   K_{A'A''}(t) = 
   -{\textstyle\frac{1}{2}}\Gamma_{A'A''}^{31}
   \Delta\omega_\mathrm{C}\,
   e^{-i(\Delta
   -i\Delta\omega_\mathrm{C})t}
\end{equation}
($\Delta$ $\!=$ $\!\omega_{\rm C}$ $\!-$ 
$\!\tilde{\omega}_{A31}$).
Using Eq.~(\ref{e53}) and differentiating both sides of 
Eq.~(\ref{e50-3}) with respect to time, we find that
$C_\pm^{13}$ satisfies the second-order differential equation
\begin{align}
\label{e54}
   \ddot{C}_\pm^{13}
   + a_{1\pm} \dot{C}_\pm^{13}
   + a_{2\pm} C_\pm^{13}=
   \dot{F}_\pm(t)
   + i(\Delta-i\Delta\omega_{\rm C})F_\pm(t),
\end{align}
where 
\begin{align}
\label{e54-1}
   a_{1\pm}
   = &\ i (\Delta\mp\Delta_{AB}^{31})
      +\Delta\omega_{\rm C}+ 
   {\textstyle\frac{1}{2}}\Gamma^{32}_{AA},
\\[1ex] 
\label{e54-2}
   a_{2\pm} = &\ 
   g_\pm^2
   +(\Delta-i\Delta\omega_{\rm C})
   \bigl(\pm\Delta_{AB}
   +i{\textstyle\frac{1}{2}}\Gamma^{32}_{AA}\bigr),
\end{align}
with
\begin{align}
\label{e54-2a}
   g_\pm^2 ={\textstyle\frac{1}{2}}\Gamma^{31}_\pm\Delta\omega_{\mathrm C},
   \quad 
   \Gamma^{31}_\pm=\Gamma^{31}_{AA}\pm\Gamma^{31}_{AB}.
\end{align}
If $C_\pm^{13}(t)$ are known, then
the probability amplitude 
$\mathbf{C}(\mathbf{r},\omega,t)$ can
be obtained 
from Eq.~(\ref{e47}) together with Eq.~(\ref{e50-1}).

To calculate the terms $\hat{\varrho}^{(n)}(t)$ 
($n$ $\!>$ $\!0$), Eq.~(\ref{e25.2}), of the series (\ref{e25}), 
we note that the action of the operator $\hat{S}$, Eq.~(\ref{e37}), 
on $\hat{\varrho}^{(0)}(t)$ $\!=$ $\!|\Psi(t)\rangle\langle\Psi(t)|$
corresponds to atomic transitions $|3\rangle$ $\!\to$ $|2\rangle$. 
Thus, only the states $|1,3\rangle$ and $|3,1\rangle$, or
equivalently $|\pm_{13}\rangle$, can contribute to
$\hat{S}[|\Psi(t)\rangle\langle\Psi(t)|]$. It is not difficult to see that
\begin{align}
\label{e56}
   &\hspace{-1ex}\hat{S}(|\pm_{13}\rangle\langle \pm_{13}|)=
\nonumber\\&
   \Gamma^{32}_{AA} |\pm_{12}\rangle\langle \pm_{12}|
   \mp{\textstyle\frac{1}{2}}\Gamma^{32}_-
   (|+_{12}\rangle\langle +_{12}|
   - |-_{12}\rangle\langle -_{12}|),
\\[1ex]
\label{e56-a}
   &\hspace{-1ex}\hat{S}(|\pm_{13}\rangle\langle \mp_{13}|)=
\nonumber\\&
   \Gamma^{32}_{AA} |\pm_{12}\rangle\langle \mp_{12}|
   -{\textstyle\frac{1}{2}}\Gamma^{32}_-
   (|\pm_{12}\rangle\langle \mp_{12}|
   - |\mp_{12}\rangle\langle \pm_{12}|)
\end{align}
[$\Gamma_\pm^{32}$ $\!=$ $\!\Gamma^{32}_{AA}$ $\!\pm$ 
$\!\Gamma^{32}_{AB}$, $|\pm_{12}\rangle$ $\!=$ 
$\!2^{-\frac{1}{2}}(|2,1\rangle$ $\!\pm$ $\!|1,2\rangle)$].
Combining Eqs.~(\ref{e55}), (\ref{e56}), and (\ref{e56-a}), we derive 
\begin{align}
\label{e56-1}
     \hat{S}&\hat{\varrho}^{(0)}(t) =
     \hat{S}[|\Psi(t)\rangle\langle\Psi(t)|] 
\nonumber\\[1ex]&
     = |\{0\}\rangle \langle \{0\}|\otimes
     \Bigl\{
     \left({\textstyle\frac{1}{2}}\Gamma_+^{32}|C_+^{13}|^2
     \right.
\nonumber\\[1ex]&\quad 
     \left.
     +{\textstyle\frac{1}{2}}\Gamma_-^{32}
     |C_-^{13}|^2\right)|+_{12}\rangle\langle +_{12}|
     +\left({\textstyle\frac{1}{2}}\Gamma_+^{32}|C_-^{13}|^2
     \right.
\nonumber\\[1ex]&\quad 
     \left.
     +{\textstyle\frac{1}{2}}\Gamma_-^{32}
     |C_+^{13}|^2\right)|-_{12}\rangle\langle -_{12}|
     +\left[
     \left({\textstyle\frac{1}{2}}\Gamma_+^{32}  C_+^{13}C_-^{13*}
     \right.
     \right.
\nonumber\\[1ex]&\quad
     \left.\left.
     +{\textstyle\frac{1}{2}}\Gamma_-^{32}
     C_+^{13*}C_-^{13}\right)|+_{12}\rangle\langle -_{12}|
     +{\rm H.c.}
     \right] \Bigr\},
\end{align}
\begin{align}
\label{e56-e} 
\hat{S}\hat{S}(|\Psi(t)\rangle\langle\Psi(t)|) = 0. 
\end{align}
Recalling that $\hat{{\cal H}}$, Eqs.~(\ref{e37.1})--(\ref{e37.3}), 
acts on atomic states in the subspace
spanned by $|\pm_{13}\rangle$, we see that  
\begin{align}
\label{e56-d}
    e^{\hat{L}(t-t_1)}
    \hat{S}(|\Psi(t_1)\rangle\langle\Psi(t_1)|) =
    \!\hat{S}(|\Psi(t_1)\rangle\langle\Psi(t_1)|),
\end{align}
leading to
\begin{eqnarray}
\label{e56-2}
   \varrho^{(1)}(t)=
   \int_0^t\di t_1\hat{S}(|\Psi(t_1)\rangle\langle\Psi(t_1)|)
\end{eqnarray}
[cf. Eq.~(\ref{e25.2})]. 
Further, Eqs.~(\ref{e56-e}) and (\ref{e56-d}) imply that
\mbox{$\hat{\varrho}^{(n)}$ $\!=$ $\!0$} if $n$ $\!\ge$ $\!2$. 
Thus, the solution to the master equation reads 
\begin{eqnarray}
\label{e57}
   \hat{\varrho}(t)=|\Psi(t)\rangle\langle\Psi(t)|
   +\int_0^t\di t_1\hat{S}[|\Psi(t_1)\rangle\langle\Psi(t_1)|]
\end{eqnarray}
together with Eqs.~(\ref{e55}) and (\ref{e56-1}).

\subsection{Stationary limit}
\label{sec3B}

Let us restrict our attention to the stationary limit
\mbox{$t$ $\!\to$ $\infty$}. Since $F_{31}(t)$ and $F_{13}(t)$
approach zero as $t$ tends to infinity, Eqs.~(\ref{e45}) and
(\ref{e46}) imply that 
\begin{align}
\label{e57-1b}
\lim_{t\to\infty}C^{13}_\pm(t) = 0.  
\end{align}
Inserting Eq.~(\ref{e55}) in Eq.~(\ref{e57})
and taking the trace with respect to the 
$\mathbf{f}$-field, we derive 
\begin{align}
\label{e57-1a}
   \mathrm{Tr}_\mathrm{field}\hat{\varrho}(t\rightarrow \infty)
   = \hat{\varrho}_{\rm at} 
,
\end{align}
\begin{eqnarray}
\label{e57-1}
\lefteqn{
    \hat{\varrho}_{\rm at} =
    \alpha_+
    |+_{12}\rangle\langle +_{12}| +
    \alpha_-
    |-_{12}\rangle\langle -_{12}|
}
\nonumber\\[1ex]&
    +\left(
    \beta
    |+_{12}\rangle\langle-_{12}| + {\rm H.c.}\right)
    +(1\!-\!\alpha_+-\alpha_-)|1,1\rangle\langle 1,1|,
\qquad
\end{eqnarray}
where
\begin{gather}
\label{e57-2}
    \alpha_\pm 
    = \int_0^\infty {\rm d}t
     \left({\textstyle\frac{1}{2}}\Gamma_\pm^{32}|C_+^{13}|^2
     +{\textstyle\frac{1}{2}}\Gamma_\mp^{32}
     |C_-^{13}|^2\right),
\\[1ex]
\label{e57-4}
     \beta
     = \int_0^\infty {\rm d}t
     \left(
     {\textstyle\frac{1}{2}}\Gamma_+^{32}  C_+^{13}C_-^{13*}
     +{\textstyle\frac{1}{2}}\Gamma_-^{32}
     C_+^{13*}C_-^{13}\right).
\end{gather}

To determine the accessible entanglement of the two
atoms, it may be instructive to study the concurrence of 
the atomic subsystem, which may be regarded as being a 
measure of entanglement \cite{wootters}. For this purpose, 
we have to calculate the spin-flipped density operator 
\begin{align}
\label{e57-4a}
     \hat{\tilde{\rho}}_{\rm at}=
     \left(\hat{\sigma}_{Ay}\otimes\hat{\sigma}_{By}\right)
     \,\hat{\rho}^*_{\rm at}\,
     \left(\hat{\sigma}_{By}\otimes\hat{\sigma}_{Ay}\right),
\end{align}
where 
\begin{align}
\label{e57-a}
\left(\sigma_{A(B)y}\right)_{mn}
\ \widehat{=}\  
\left(
\begin{array}{rr}
0 & -i\\i & 0
\end{array}
\right)  
\end{align} 
[$m(n)$ $\!=$ $\!1,2$],
and to determine the two nonzero eigenvalues $\lambda_\pm$ of 
$\hat{\rho}_{\rm at}\hat{\tilde{\rho}}_{\rm at}$. 
A somewhat lengthy but straightforward calculation yields
\begin{eqnarray}
\label{e57-4b}
\lefteqn{
    \lambda_\pm={\textstyle\frac{1}{2}}
    \left\{\alpha_+^2+\alpha_-^2
    -2\left[({\rm Re}\,\beta)^2-({\rm Im}\,\beta)^2\right] \right\}
}
\nonumber\\[1ex]
    &\pm\,\frac{1}{2} 
    \sqrt{ 
    \left[(\alpha_+\!+\!\alpha_-)^2 -4 ({\rm Re}\beta)^2 \right]
    \left[(\alpha_+\!-\!\alpha_-)^2 +4 ({\rm Im}\beta)^2 \right]
    }\,, 
\nonumber\\
\end{eqnarray}
which then determine the concurrence  
\begin{eqnarray}
\label{e57-4c}
     {\cal C}=
     \sqrt{\lambda_+}-\sqrt{\lambda_-}\,,
\end{eqnarray}
the value of which is in the interval $[0,1]$.
The nearer 1 the value of $\mathcal{C}$
is, the higher is the degree of entanglement.
Equations (\ref{e57-4b}) and (\ref{e57-4c})
reveal that a noticeably entangled state of the two atoms
can be generated if
\begin{equation}
\label{e57-4d}
\alpha_+(\alpha_-) \gg \alpha_-(\alpha_+),|\beta|,  
\end{equation}
thus ${\cal C} \rightarrow \alpha_+(\alpha_-)$.
Needless to say that the entanglement condition (\ref{e57-4d})
is already expected from inspection of Eq.~(\ref{e57-1}).

\subsection{Different Coupling Regimes}
\label{sec3C}

Let us return to Eq.~(\ref{e54}) and focus on the case where
\begin{equation}
\label{e54-0}
   \dot{F}_\pm(t) \simeq 
   - i(\Delta-i\Delta\omega_\mathrm{C})F_\pm(t)
\end{equation}
is valid, so that the term on the right-hand side in 
Eq.~(\ref{e54}) can be omitted. Obviously, this is the case 
when initially the (Lorentzian) field resonance of
mid-frequency $\omega_\mathrm{C}$ and width 
$\Delta\omega_\mathrm{C}$ is excited (for details, 
see Sec.~\ref{sec3D}). Under the initial conditions
\begin{equation}
\label{e54-3}
    C_\pm^{13}(0)=0,\qquad
    \dot{C}_\pm^{13}(0)=F_\pm(0),
\end{equation}
the solution to Eq.~(\ref{e54}) can then be written in the form of
\begin{equation}
\label{e54-4}
    C_\pm^{13}(t) = \frac{F_\pm(0)}{q_\pm}
    e^{-a_{1\pm}t/2}
    (e^{q_\pm t/2} - e^{-q_\pm t/2}),
\end{equation}
where
\begin{equation}
\label{e54-5}
    q_\pm=\sqrt{a_{1\pm}^2 - 4 a_{2\pm}}\ .
\end{equation}
Restricting again our attention to the stationary limit,
we further assume, for simplicity, both the detuning
$\Delta$ and the dipole-dipole coupling strength 
$\Delta_{AB}^{31}$ vanish, i.e., \mbox{$\Delta$ $\!=$ $\!0$} 
and $\Delta_{AB}^{31}$ $\!=$ $\!0$. Since even under these
conditions the explicit form of the expansion coefficients 
$\alpha_\pm$, Eq.~(\ref{e57-2}), and $\beta$, Eq.~(\ref{e57-4}),
is rather involved, we renounce its presentation here but
consider instead some instructive special cases. 

{F}rom Eqs.~(\ref{e54-1}) and (\ref{e54-4}) it is seen that 
the damping constant of $C_\pm^{13}$ is determined by the sum of the 
half width at half maximum of the field resonance strongly 
coupled to the transition $|3\rangle$ $\!\leftrightarrow$ 
$\!|1\rangle$ and the half width at half maximum of
the transition $|3\rangle$ $\!\to$ $\!|2\rangle$,
$\Delta\omega_\mathrm{C}$ and $\Gamma^{32}_{AA}/2$,
respectively. Due to the finite $\Delta\omega_\mathrm{C}$, 
an atom tends to occupy the state $|1\rangle$, while the 
effect of the finite $\Gamma^{32}_{AA}$ 
is that the atom prefers to occupy the state $|2\rangle$.
We may therefore restrict ourselves to situations in which
\begin{equation}
\label{e57-5}
      \Gamma_{AA}^{32} \gg \Delta\omega_C .
\end{equation}
To achieve noticeable entanglement, the interatomic coupling 
should be sufficiently strong, i.e., 
$|\Gamma_{AB}^{31}|$ $\!\to$ $\!\Gamma_{AA}^{31}$ and 
$|\Gamma_{AB}^{32}$ $\!|\to$ $\!\Gamma_{AA}^{32}$,
equivalently, 
\begin{align}
\label{e57-6a}
\frac{\Gamma_\pm^{31}}{\Gamma_\mp^{31}} \gg 1,
\quad
\frac{\Gamma_{\pm(\mp)}^{32}}{\Gamma_{\mp(\pm)}^{32}} \gg 1.
\end{align}
Note that the first inequality is equivalent to 
$g_\pm$ $\!\gg$ $\!g_\mp$ [cf. Eq.~(\ref{e54-2a})].
We now distinguish between the following three cases.

\subsubsection*{}
\vspace{-5ex}
\centerline{ (a) \quad 
$g_\pm\gg\Gamma^{32}_{AA} \gg
\Delta\omega_{\rm C}\gg g_\mp$
}
\vspace{2ex}

In this case, either the symmetric state $|+_{13}\rangle$
or the antisymmetric state $|-_{13}\rangle$ is strongly 
coupled to the medium-assisted electromagnetic field
whereas the other one is weakly coupled.
For the strongly and weakly-coupled states, 
respectively, Eq.~(\ref{e54-4}) approximates to
\begin{align}
\label{e58-2}
   C^{13}_\pm(t)=\frac{F_\pm(0)}{g_\pm}\,
    e^{-\Gamma^{32}_{AA}t/4}\sin (g_\pm t) ,
\end{align}
and 
\begin{align}
\label{e58-3} 
   C^{13}_\mp(t)= \frac{2F_\mp(0)}{\Gamma^{32}_{AA}}\,
   \left[e^{-\Delta\omega_{C} t}
   -e^{-\Gamma^{32}_{AA}t/2}\right].
\end{align}
It is seen that $C^{13}_\pm(t)$ undergoes damped Rabi oscillations 
of frequency $g_\pm$, while $C^{13}_\mp(t)$
undergoes a two-channel exponential decay.
The steady-state density operator parameters $\alpha_\pm$, 
Eq.~(\ref{e57-2}), and $\beta$, Eq.~(\ref{e57-4}), approximate to
\begin{gather}
\label{e59-3}
    \alpha_\pm = {\textstyle\frac{1}{2}}\Gamma_{\pm(\mp)}^{32}\,
    \frac{|F_{+(-)}(0)|^2}{g_{+(-)}^2 \Gamma^{32}_{AA}}
    + \Gamma_{\mp(\pm)}^{32}\,\frac{|F_{-(+)}(0)|^2}
    {(\Gamma^{32}_{AA})^2 \Delta \omega_C}\,,
\\[1ex]
\label{e59-5}
     \beta 
     = \left[\Gamma_+^{32}F_+(0)F_-^\ast(0)
     + \Gamma_-^{32}F_+^\ast(0)F_-(0)
     \right]
    \frac{\Gamma^{32}_{AA}}{2g_{+(-)}^4}
\end{gather}
for $g_{+(-)}$ $\!\gg$ $\!g_{-(+)}$.

\subsubsection*{}
\vspace{-5ex}
\centerline{ (b) \quad $g_\pm \gg g_\mp 
\gg \Gamma^{32}_{AA} \gg\Delta\omega_C $
} 
\vspace{2ex}

When both $g_\pm$ and $g_\mp$ dominate the other parameters, 
then the states $|+_{13}\rangle$ and $|-_{13}\rangle$ are 
both strongly coupled to the medium-assisted electromagnetic field, 
and Eq.~(\ref{e54-4}) approximates to
\begin{equation}
\label{e61-1}
   C^{13}_\pm(t)=\frac{F_\pm(0)}{g_\pm}\,
   e^{-\Gamma^{32}_{AA}t/4}\sin (g_\pm t), 
\end{equation}
which is exactly analogous to Eq.~(\ref{e58-2}).
The steady-state density operator parameters 
$\alpha_\pm$ and $\beta$ take the approximate form of
\begin{equation}
\label{e61-3}
    \alpha_\pm = {\textstyle\frac{1}{2}}\Gamma_{\pm}^{32}\,
    \frac{|F_{+}(0)|^2}{g_{+}^2 \Gamma^{32}_{AA}}
    +{\textstyle\frac{1}{2}}
    \Gamma_{\mp}^{32}\,\frac{|F_{-}(0)|^2}
    {g_-^2 \Gamma^{32}_{AA}}
\end{equation}
and, for $g_{+(-)}$ $\!\gg$ $\!g_{-(+)}$,
\begin{equation}
\label{e61-5}
     \beta 
     = \left[\Gamma_+^{32}F_+(0)F_-^\ast(0)
     + \Gamma_-^{32}F_+^\ast(0)F_-(0)
     \right]
    \frac{\Gamma^{32}_{AA}}{2g_{+(-)}^4}\,.
\end{equation}

\subsubsection*{}
\vspace{-5ex}
\centerline{ (c) \quad $\Gamma^{32}_{AA} \gg g_\pm 
\gg g_\mp,\,\Delta\omega_C $
}
\vspace{2ex}

When the value of $\Gamma^{32}_{AA}$ sufficiently 
exceeds the values
of the other parameters, then from Eq.~(\ref{e54-4}) it follows that
\begin{equation}
\label{e63-1} 
   C^{13}_\pm(t)= \frac{2F_\pm(0)}{\Gamma^{32}_{AA}}
   \left[e^{-\Delta\omega_{C} t}
   -e^{-\Gamma^{32}_{AA}t/2}\right],
\end{equation} 
i.e., the behavior typical of weakly-coupled states 
is observed [cf. Eq.~(\ref{e58-3})]. In this approximation, 
the steady-state density operator parameters 
$\alpha_\pm$ and $\beta$ read
\begin{align}
\label{e64}
   \alpha_\pm = &\; \Gamma_\pm^{32}\,
   \frac{|F_+(0)|^2}
   {(\Gamma_{AA}^{32})^2
   (\Delta\omega_{\rm C} + 2g_+^2/\Gamma_{AA}^{32})}
\nonumber\\[1ex]
   &+ \Gamma_\mp^{32}\,
   \frac{|F_-(0)|^2}
   {(\Gamma_{AA}^{32})^2
   (\Delta\omega_{\rm C} + 2g_-^2/\Gamma_{AA}^{32})}
\end{align}
and, for $g_{+(-)}$ $\!\gg$ $\!g_{-(+)}$, 
\begin{equation}
\label{e64-2}
    \beta = \frac{\Gamma_+^{32}F_+(0)F_-^\ast(0)
     + \Gamma_-^{32}F_+^\ast(0)F_-(0) }
    {(\Gamma^{32}_{AA})^2
   (\Delta\omega_{\rm C} + g_{+(-)}^2/\Gamma_{AA}^{32})}.
\end{equation}

\subsection{Preparation of the initial state}
\label{sec3D}

One possible way to initially prepare the medium-assisted 
electromagnetic field in the desired quantum state (\ref{e38}),   
is to use an additional atom, say atom $D$, such that 
$\tilde{\omega}_{D31}$ $\!=$ $\!\tilde{\omega}_{A31}$ $\!=$ 
$\!\tilde{\omega}_{B31}$ $\!=$ $\!\omega_C$. Let the transition 
\mbox{$|1\rangle$ $\!\leftrightarrow$ $\!|3\rangle$} of atom $D$ 
strongly interact with the medium-assisted electromagnetic field 
in the absence of atoms $A$ and $B$. This can be achieved, 
for instance, by using atomic beams and letting 
atom $D$ pass the equipment before atoms $A$ and $B$ pass it. 
When atom $D$ initially prepared in the excited state $|3\rangle$
strongly interacts with the medium-assisted electromagnetic field
initially prepared in the vacuum state, then an interaction 
time can be chosen after which the atomic excitation 
is transferred to the field.

The probability amplitude of finding, 
after some interaction time $\Delta t$, atom $D$ (regarded as 
an effective two-level system) in the ground state 
and the $\hat{\mathbf{f}}$-field
in a single-quantum state is \cite{Ho02}
\begin{eqnarray}
\label{e67}
\lefteqn{
   {\bf C}(\rb,\omega,t=0)
}
\nonumber\\ &&\hspace{-2ex}
   = \frac{i}{\hbar}
   \!\int_{-\Delta t}^0 \!\!\di t'\, 
   \db_{D31}^*\tilde{\bm{G}}^*(\rb_D,\rb,\omega)
   e^{i(\omega-\tilde{\omega}_{D31})t'}C_{U_D}(t'),
\qquad
\end{eqnarray}
where 
\begin{equation}
\label{e67-1}
    C_{U_D}(t)=
    e^{-\Delta\omega_{\rm C}(t+\Delta t)/2}
    \cos [g_D(t+\Delta t)]
\end{equation}
is the probability amplitude of finding the atom in the upper state. Here,
\begin{equation}
\label{e67-1a}
    g_D=
   \sqrt{\Gamma_{DD}^{31}\Delta\omega_{\rm C}/2}
\end{equation}
is the single-atom Rabi frequency, with 
$\Gamma_{DD}^{31}$ being determined according to Eq.~(\ref{e52}). 
Substitution of Eq.~(\ref{e67}) into Eqs.~(\ref{e50}) and 
(\ref{e50.1}) yields
\begin{eqnarray}
\label{e67-2}
    F_{31}(t) = \int_{-\Delta t}^0 {\rm d} t'\,
    K_{AD}(t-t') C_{U_D}(t'),
\\[1ex]
\label{e67-3}
    F_{13}(t) = \int_{-\Delta t}^0 {\rm d} t'\,
    K_{BD}(t-t') C_{U_D}(t'),
\end{eqnarray}
where $K_{BD}(t)$ is defined according to Eq.~(\ref{e49}).
Note that $F_\pm(t)$, Eq.~(\ref{e50-5}), calculated by using
$F_{31}$ and $F_{13}$ given in Eqs.~(\ref{e67-2}) and 
(\ref{e67-3}) fulfills Eq.~(\ref{e54-0}). To calculate $F_\pm(0)$, 
we fix the interaction time $\Delta t$ such that 
\mbox{$C_{U_D}(0)$ $\!=$ $\!0$}, thus
\begin{equation}
\label{e67-4}
    \Delta t = \frac{\pi}{2 g_D}\,.
\end{equation}
Combining Eq.~(\ref{e50-5}) with Eqs.~(\ref{e67-1})--(\ref{e67-4}),
we derive, on applying the Lorentz approximation according 
to Eq.~(\ref{e51}),
\begin{equation}
\label{e67-5}
     F_\pm(0) = -\frac{1}{\sqrt{2}} 
     \frac{g_{D\pm}^2}{g_D}
     \exp\!\left(-\Delta \omega_C \frac{\pi}{2g_D}\right),
\end{equation}
where
\begin{equation}
\label{e67-6}
     g_{D\pm} = 
     \sqrt{(\Gamma^{31}_{BD} \pm \Gamma^{31}_{AD})
     \Delta \omega_C/2}\,,
\end{equation}
and $\Gamma^{31}_{AD}$ and $\Gamma^{31}_{BD}$ are 
defined according to Eq.~(\ref{e52}). 

In Eq.~(\ref{e67-5}), the exponential factor characterizes
the photon loss during the interaction time due to the finite 
width of the field resonance. Obviously, the better the strong-coupling
condition $\Delta \omega_C$ $\!\ll$ $\!g_D$ is fulfilled, the less
the photon loss is. In particular, when atom $A$ (or $B$) 
changes places with atom $D$ and the orientations 
of the transition dipole moments of atoms $A$ (or $B$)
and $D$ are the same, then from Eq.~(\ref{e67-5}) it follows that 
\mbox{($\Delta\omega_C$ $\!\ll$ $\!g_D$)} 
\begin{equation}
\label{e72-0}
  F_\pm(0) \simeq 
  -g_\pm, \quad  F_\mp(0) \simeq -g_\mp^2/g_\pm. 
\end{equation}
It is worth noting that, as we will see in Sec.~\ref{sec5}, 
the highest degree of entanglement can be achieved in case 
of equal positions of atoms $D$ and $A$ (or $B$). 

\section{ Atomic entanglement near a dielectric microsphere}
\label{sec5}

Let us apply the theory to two atoms near a dispersing and
absorbing dielectric microsphere (of radius $R$)  
characterized by a Drude-Lorentz type permittivity
\begin{equation}
\label{e69}
   \varepsilon(\omega)
   =1+\frac{\omega_\mathrm{P}^2}
   {\omega_\mathrm{T}^2-\omega^2-i\omega\gamma}
\end{equation}
($\omega_\mathrm{P}$, coupling constant; $\omega_\mathrm{T}$, 
transverse resonance frequency; $\gamma$, absorption parameter), 
which features a band gap in the region 
$\omega_\mathrm{T}$ $\!<$ $\!\omega$ $\!<\omega_\mathrm{L}$ $\!=$ 
$\!\sqrt{\omega_\mathrm{T}^2+\omega_\mathrm{P}^2}$, 
where \mbox{$\mathrm{Re}\,\varepsilon(\omega)$ $\!<$ $\!0$}.

\subsection{Two-atom coupling}
\label{sec5.1}

Making use of the Green tensor for a dielectric sphere \cite{leong},
one can show, on assuming radial dipole orientations,  
that Eq.~(\ref{e9}) leads to 
\begin{align}
\label{e70}
   &\Gamma_{A'A''} \equiv \Gamma_{A'A''}^{mn} = 
   {\textstyle\frac{3}{2}}\Gamma_0
   {\rm Re}\!\sum_{l=1}^\infty 
   \frac{l(l+1)(2l+1)}{(kr)^2}\,h_l^{(1)}(kr)
\nonumber\\[1ex]&\hspace{2ex}\times\,
   \left[j_l(kr)+B_l^N(\omega)h_l^{(1)}(kr)\right]
   P_l(\cos\theta)
\end{align}
[$\omega$ $\!\equiv$ $\!\tilde\omega_{A'}^{mn}$ 
$\!=$ $\!\tilde\omega_{A''}^{mn}$ $\!>$ $\!0$; $k$ $\!=$ $\!\omega/c$;
$r$ $\!\equiv$ $\!r_{A'}$ $\!=$ $\!r_{A''}$ $\!(>$ $\!R)$, 
radial position of the atoms]. Here, $\Gamma_0$ is the single-atom 
decay rate in free space, $j_l(z)$ and $h^{(1)}_l(z)$ are the 
spherical Bessel and Hankel functions, respectively, 
$P_l(x)$ is the Legendre function, $\theta$ is the angle between 
the two transition dipole moments
($|\mathbf{d}_{A'mn}|$ $\!=$ $\!|\mathbf{d}_{A''mn}|$), 
and the scattering coefficients $B_l^N(\omega)$ read \cite{leong}
\begin{equation}
\label{e71}
   B_l^N(\omega)
   =-\frac{\varepsilon(\omega)j_l(z_2)[z_1j_l(z_1)]'
   -j_l(z_1)[z_2j_l(z_2)]'}
   {\varepsilon(\omega)j_l(z_2)[z_1h^{(1)}_l(z_1)]'
   -h^{(1)}_l(z_1)[z_2j_l(z_2)]'}\,,
\end{equation}
where $z_i$ $\!=$ $\!k_iR$, $k_1$ $\!=$ $\!k$, 
and $k_2$ $\!=$ $\!\sqrt{\varepsilon(\omega)}\,\omega/c$.
Note that radially oriented dipoles couple only to TM waves, 
whereas tangentially oriented dipoles couple to both 
TM and TE waves (for details, see, e.g., \cite{dkw1}).
Needless to say that \mbox{$\theta$ $\!=$ $\!0$} in case of 
a single atom ($A'$ $\!=$ $\!A''$). 

The complex roots of the denominator of the reflection coefficients 
$B_l^N(\omega)$ determine the positions and the widths of the 
sphere-assisted electromagnetic field resonances. 
When $\omega$ coincides with a resonance frequency, 
say $\omega_C$, then the corresponding 
$l$ term in Eq.~(\ref{e70}) is the leading one, thus
\begin{align}
\label{e72}
   &\Gamma_{A'A''}^{mn} \simeq 
   {\textstyle\frac{3}{2}}\Gamma_0
   {\rm Re}\biggl\{ 
   \frac{l(l+1)(2l+1)}{(kr)^2}\,h_l^{(1)}(kr)
\nonumber\\[1ex]&\hspace{2ex}\times\,
   \left[j_l(kr)+B_l^N(\omega)h_l^{(1)}(kr)\right]
   P_l(\cos\theta)\biggr\}
\end{align}
($\omega$ $\!\simeq$ $\!\omega_C$). 
Equation (\ref{e72}) implies that
when the two atoms ($A'$ $\!\neq$ $\!A''$) are 
at diametrically opposite 
positions with respect to the sphere, i.e., 
$\theta$ $\!=$ $\!\pi$ and hence $P_l(\cos\theta)$ $\!=$ $\!(-1)^l$,
then the interaction of the symmetric (antisymmetric) 
state with the sphere-assisted electromagnetic field
is enhanced, while the antisymmetric (symmetric) state
almost decouples [cf. Eq.~(\ref{e54-2a})]. 

\begin{figure}
\noindent

\begin{center}
\includegraphics[width=.9\linewidth]{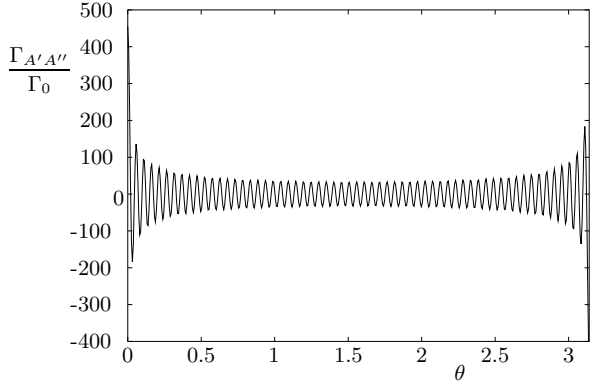}
\end{center}
\caption{%
The two-atom collective decay rate  
$\Gamma_{AA'}^{mn}$ [Eq.~(\ref{e70}), \mbox{$A'$ $\!\neq$ $\!A''$}]
as a function of the angle $\theta$ between the transition dipole 
moments for $\omega$ $\!=$ $\!1.0501\,\omega_\mathrm{T}$.
The two atoms are at distances $\Delta r$ $\!\equiv$ 
\mbox{$\!r$ $\!-$ $\!R$} $\!=$ $\!0.14\lambda_\mathrm{T}$ 
$(\lambda_\mathrm{T}$ $\!=$ $\!2\pi c/\omega_\mathrm{T})$
from the surface of a dielectric sphere 
($\omega_\mathrm{P}$ $\!=$ $\!0.5\,\omega_\mathrm{T}$, 
$\gamma$ $\!=$ $\!10^{-6}\omega_\mathrm{T}$, 
$R$ $\!=$ $\!10\,\lambda_\mathrm{T}$).}
\label{fig2}
\end{figure}%
The dependence on $\theta$ of $\Gamma_{A'A''}$ ($A'$ $\!\neq$
$\!A''$) as given by Eq.~(\ref{e70}) is illustrated 
in Fig.~\ref{fig2}, where the atomic transition 
frequency $\omega$ is chosen to be
close to a microsphere resonance frequency. 
\begin{figure}

\noindent
\begin{center}
\includegraphics[width=.9\linewidth]{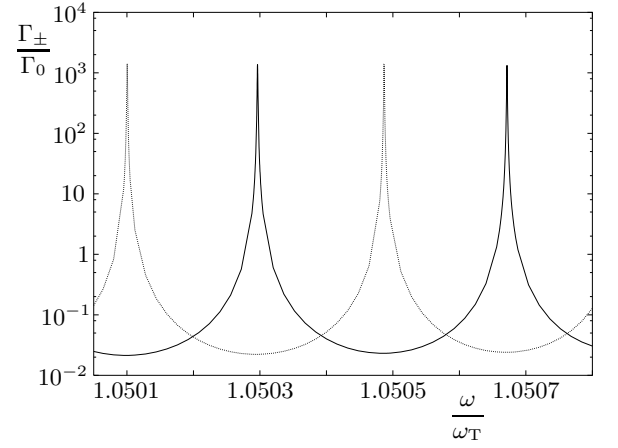}
\end{center}
\caption{%
The two-atom decay rates 
$\Gamma_+$ $\!=$ $\!\Gamma_{A'A'}$ $\!+$ $\!\Gamma_{A'A''}$
(solid curve) and 
$\Gamma_-$ $\!=$ $\!\Gamma_{A'A'}$ $\!-$ $\!\Gamma_{A'A''}$
(dotted curve) for the symmetric and antisymmetric states,
respectively,
as functions of the transition frequency $\omega$, with 
$\Gamma_{A'A''}$ from Eq.~(\ref{e70}) for \mbox{$\theta$ $\!=$ $\!\pi$}.
The other parameters are the same as in Fig.~\ref{fig2}].
}
\label{fig3}
\end{figure}%
From Figs.~\ref{fig3} and \ref{fig5} it is clearly seen 
that the value of $\Gamma_+$ ($\Gamma_-$) can drastically 
exceed the value of $\Gamma_-$ ($\Gamma_+$) when the two atoms 
approach the microsphere and the transition frequency equals
a resonance frequency. Recall that $\Gamma_+$ ($\Gamma_-$) 
is a measure of the strength of coupling of the
symmetric (antisymmetric) state to the sphere-assisted field. 
\begin{figure}

\noindent
\begin{center}
\includegraphics[width=.9\linewidth]{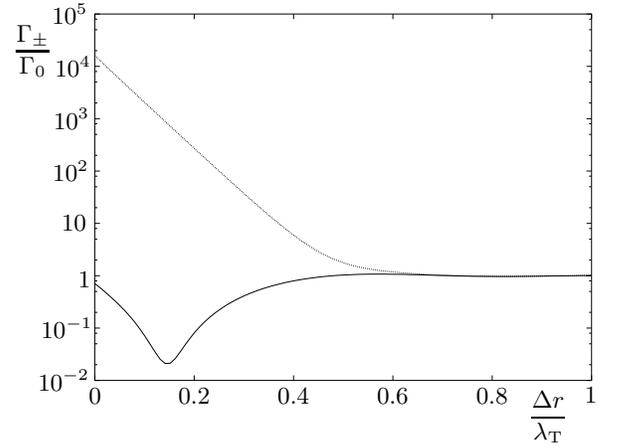}
\end{center}
\caption{%
The two-atom decay rates 
$\Gamma_+$ $\!=$ $\!\Gamma_{A'A'}$ $\!+$ $\!\Gamma_{A'A''}$
(solid curve) and 
$\Gamma_-$ $\!=$ $\!\Gamma_{A'A'}$ $\!-$ $\!\Gamma_{A'A''}$
(dotted curve) for the symmetric and antisymmetric states,
respectively, as functions of the 
distance $\Delta r$ between the atom and the surface 
of the dielectric sphere,   
with $\Gamma_{A'A''}$ from Eq.~(\ref{e70}) 
for \mbox{$\theta$ $\!=$ $\!\pi$}.
The other parameters are the same as in Fig.~\ref{fig2}].
}
\label{fig5}
\end{figure}%
In particular, Fig.~\ref{fig5} reveals that there is an optimum
distance -- the distance at which the solid curve attains
the minimum -- for which the best contrast between 
$\Gamma_+$ and $\Gamma_-$ can be realized. 
With increasing distance of the atoms from the sphere,
the values of both $\Gamma_+$ and $\Gamma_-$
tend to the free-space value $\Gamma_0$ as they should. 

\begin{figure}

\noindent
\begin{center}
\includegraphics[width=.9\linewidth]{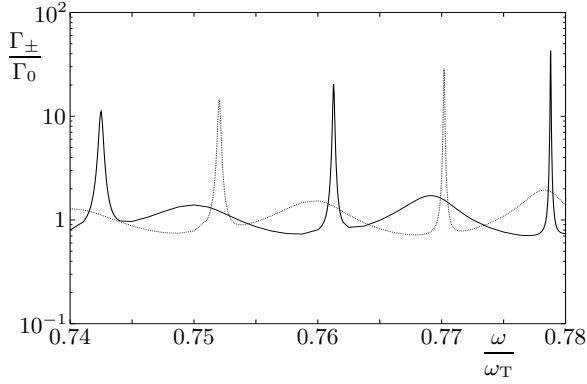}
\end{center}
\caption{%
The two-atom decay rates 
$\Gamma_+$ $\!=$ $\!\Gamma_{A'A'}$ $\!+$ $\!\Gamma_{A'A''}$
(solid curve) and 
$\Gamma_-$ $\!=$ $\!\Gamma_{A'A'}$ $\!-$ $\!\Gamma_{A'A''}$
(dotted curve) for the symmetric and antisymmetric states,
respectively,
as functions of the transition frequency $\omega$, with 
$\Gamma_{A'A''}$ from Eq.~(\ref{e70}) for \mbox{$\theta$ $\!=$ $\!\pi$}.
The other parameters are the same as in Fig.~\ref{fig2}].
}
\label{fig4}
\end{figure}

Figures \ref{fig2}--\ref{fig5} refer to atomic transition 
frequencies within the band gap. In this case, the strong two-atom 
interaction observed when the atoms are
at diametrically opposite positions with respect to the sphere    
is mediated by SG waves.
Of course, the effect of enhanced $\Gamma_+$ ($\Gamma_-$)
and simultaneously reduced $\Gamma_-$ ($\Gamma_+$)
can also be observed for transition frequencies
below the band gap. In this case, the cavity-assisted
field resonances correspond to WG waves. An example  
is shown in Fig.~\ref{fig4}. 
Figures \ref{fig3} and \ref{fig4} also convey a feeling of
the sharpness of the field resonances,
which ranges from being very sharp to being less so.
The sharpness can be improved by increasing the microsphere 
radius or by reducing the material absorption. Note that WG waves
much more suffer from absorption than do SG waves
(see, e.g., Ref.~\cite{dkw1}).

\subsection{Entanglement of two $\mathbf{\Lambda}$-type atoms}
\label{sec5.2}

The results given in Sec.~\ref{sec5.1} show that 
the optimal positions of two $\Lambda$-type atoms $A$ and $B$,
which are desired to entangle with each other near a microsphere,
are diametrically opposite with respect to the sphere.
Further, the transition frequency $\tilde{\omega}_{A31}$ $\!=$
$\!\tilde{\omega}_{B31}$
should coincide with the (mid-)frequency $\omega_C$ of a sufficiently
sharply peaked sphere-assisted field resonance, so that the 
strong-coupling regime is realized and the first of the conditions 
(\ref{e57-6a}) is satisfied. Finally, the 
transition frequency $\tilde{\omega}_{A32}$ $\!=$
$\!\tilde{\omega}_{B32}$ should coincide with the (mid-)frequency 
of some moderately peaked sphere-assisted field resonance, 
so that the second of the conditions (\ref{e57-6a}) is also satisfied,
but the weak-coupling regime applies, thereby giving rise to an 
irreversible decay channel. As a result, the condition (\ref{e57-5}) 
can also be expected to be satisfied.  
By choosing atoms with appropriate 
transition dipole matrix elements well matching 
cavity-assisted field resonances (for more detailed estimations, 
see Ref.~\cite{Ho02}), all the conditions
including both the inequalities characterizing the 
three cases (a)--(c) in Sec.~\ref{sec3C}
and the field-preparation conditions (\ref{e72-0})
can be fulfilled. Let us examine the cases (a)--(c) 
in more detail.

\subsubsection*{}
\vspace{-5ex}
\centerline{
(a)
\quad 
$g_\pm \gg\Gamma^{32}_{AA}
\gg\Delta\omega_{\rm C}\gg g_\mp$
} 
\vspace{2ex}

For definiteness, let  
$\Gamma_+^{31}$ $\!\gg$ $\!\Gamma_-^{31}$
and \mbox{$\Gamma_+^{32}$ $\!\gg$ $\!\Gamma_-^{32}$}.
When atom $A$ (or $B$) changes places with atom $D$, which provides 
the initial field excitation, and Eq.~(\ref{e72-0}) applies, then
Eqs.~(\ref{e59-3}) and (\ref{e59-5}) lead to
\begin{equation}
\label{e72-1}
    \alpha_+ \simeq 1,  
\end{equation}
\begin{equation}
\label{e72-2}
    \alpha_- \simeq
     \frac{\Gamma_-^{32}}{2\Gamma_{AA}^{32}}
     + \frac{2g_-^4} {g_+^2  \Gamma_{AA}^{32} \Delta \omega_C}
     \ll 1,  
\end{equation}
\begin{equation}
\label{e72-3}
    \beta \simeq  
    \left( \frac{\Gamma_{AA}^{32}g_-}
    {g_+^2}\right)^2 \ll 1.
  \end{equation}
Hence, an almost perfectly entangled state is produced, 
$\hat{\varrho}_\mathrm{at}$ $\!\simeq$
$\!|+_{12}\rangle\langle+_{12}|$ [see Eq.~(\ref{e57-1})], 
and, accordingly, $\mathcal{C}$ $\!\simeq$ $\!1$ is achieved.
Clearly, $\alpha_+$ $\!=$ $\!1$ ($\mathcal{C}$ $\!=$ $\!1$) 
cannot be exactly realized, because of the losses unavoidably 
associated with the always finite width of the field resonance. 
It is worth mentioning that when the positions of atoms $D$
and $A$ (or $B$) are different from each other
(e.g., atom $D$ was equidistant from atoms $A$ and $B$),
then the degree of entanglement that can be achieved 
is smaller than that in case of equal positions in general.     
Note that when  
\mbox{$\Gamma_-^{31}$ $\!\gg$ $\!\Gamma_+^{31}$} and
$\Gamma_-^{32}$ $\!\gg$ $\!\Gamma_+^{32}$, then 
$\hat{\varrho}_\mathrm{at}$ $\!\simeq$
$\!|+_{12}\rangle\langle+_{12}|$ is also valid.
For
\mbox{$\Gamma_\pm^{31}$ $\!\gg$ $\!\Gamma_\mp^{31}$} 
and
$\Gamma_\mp^{32}$ $\!\gg$ $\!\Gamma_\pm^{32}$, 
however, the roles of $\alpha_+$ and $\alpha_-$
are interchanged and $\hat{\varrho}_\mathrm{at}$ $\!\simeq$
$\!|-_{12}\rangle\langle-_{12}|$. 

In the scheme, the two-atom system undergoes, e.g., fast 
$|1,1\rangle$ $\!\leftrightarrow$ $|+_{13}\rangle$
Rabi oscillations as long as one of the two atoms 
jumps to state $|2\rangle$, but we do not know 
which one. Hence, the result is the entangled
state between one atom in the state $|2\rangle$
and the other in the state $|1\rangle$.
The time after which the stationary limit 
is established is determined by the lifetime
$\sim(\Gamma_{AA}^{32})^{-1}$ of the short-living
state $|+_{13}\rangle$, while the long-living
state $|-_{13}\rangle$ of lifetime $\sim(\Delta\omega_C)^{-1}$  
is practically unpopulated [cf. Eqs.~(\ref{e58-2}) 
and (\ref{e58-3})].

\subsubsection*{}
\vspace{-5ex}
\centerline{ (b) \quad 
$g_\pm \gg g_\mp \gg \Gamma^{32}_{AA}
 \gg \Delta\omega_{\rm C}$
} 
\vspace{2ex}

For definiteness, we again assume that 
$\Gamma_+^{31}$ $\!\gg$ $\!\Gamma_-^{31}$
and \mbox{$\Gamma_+^{32}$ $\!\gg$ $\!\Gamma_-^{32}$}. {F}rom 
Eqs.~(\ref{e61-3}) and (\ref{e61-5}) together with Eq.~(\ref{e72-0})
we obtain
\begin{equation}
\label{e72-4}
    \alpha_+ \simeq 1,
\end{equation}
\begin{equation}
\label{e72-5}
     \alpha_- \simeq 
     \frac{\Gamma_-^{32}}
        {2\Gamma_{AA}^{32}}
     + \frac{2g_-^2} {g_+^2}
     \ll 1,
\end{equation}
\begin{equation}
\label{e72-6}
     \beta \simeq 
     \left( \frac{\Gamma_{AA}^{32}g_-}
     {g_+^2}\right)^2
     \ll 1.
\end{equation}
Thus, this coupling regime leaves the two atoms 
in an entangled state analogous to case (a).
However, since the inequality $g_\pm$ $\!\gg$ $\!g_\mp$
requires $g_\pm$ to be as large as possible and 
$g_\mp$ to be as small as possible, while the 
inequality \mbox{$g_\mp$ $\!\gg$ $\!\Gamma^{32}_{AA}$}  
requires that $g_\mp$ is much lager than $\Gamma^{32}_{AA}$, 
it may be more difficult to realize this regime.
Note that 
$g_\mp$ is the smallest or one of the smallest parameters
in the cases (a) and (c).

\subsubsection*{}
\vspace{-5ex}
\centerline{ (c) \quad $\Gamma^{32}_{AA}
\gg g_\pm \gg g_\mp, \Delta\omega_{\rm C}$
} 
\vspace{2ex}

In this case, the irreversible decay 
from state $|3\rangle$ to state $|2\rangle$ is so dominant 
that Rabi oscillations are fully suppressed 
in the time evolution of both $C_+^{13}$ and $C_-^{13}$ 
[see Eq.~(\ref{e63-1})]. {F}rom Eq.~(\ref{e64}) we obtain, 
on again assuming $\Gamma_+^{31}$ $\!\gg$ $\!\Gamma_-^{31}$
and $\Gamma_+^{32}$ $\!\gg$ $\!\Gamma_-^{32}$
and making use of Eq.~(\ref{e72-0}),
\begin{equation}
\label{e73}
   \alpha_+ \simeq 
   \frac{2g_+^2/\Gamma_{AA}^{32}}
   {\Delta \omega_C + 2g_+^2/\Gamma_{AA}^{32}}\,.
\end{equation}
To generate the entangled state $|+_{12}\rangle$, 
i.e., $\alpha_+$
$\!\simeq$ $\!1$, the additional condition
\begin{equation}
\label{e74}
   \frac{g_+}{\Gamma_{AA}^{32}} \gg
   \frac{\Delta \omega_C}{g_+}
\end{equation}
must be required to be satisfied, as can be seen from 
Eq.~(\ref{e73}). The parameters $\alpha_-$ and $\beta$ then read
\begin{equation}
\label{e74-1}
     \alpha_- \simeq
     \frac{\Gamma_-^{32}}
        {2\Gamma_{AA}^{32}}
     + \frac{g_-^2} {g_+^2}
     \ll 1,
\end{equation}
\begin{equation}
\label{e74-2}
     \beta \simeq 
     \frac{2g_-^2}{g_+^2}
     \ll 1.
\end{equation}
In a similar fashion, it can be shown that 
in case of \mbox{$\Gamma_\pm^{31}$ $\!\gg$ $\!\Gamma_\mp^{31}$}
and $\Gamma_\mp^{32}$ $\!\gg$ $\!\Gamma_\pm^{32}$ 
the antisymmetric entangled state $|-_{12}\rangle$ is generated.

The inequality (\ref{e74}) can be understood as follows.
For $F_+(0)$ $\!\simeq -g_+$, Eq.~(\ref{e63-1}) yields
\begin{equation}
\label{e74-3} 
      C^{13}_+(t) \simeq -(2g_+/\Gamma^{32}_{AA})  
      \left[e^{-\Delta\omega_{C} t}
      -e^{-\Gamma^{32}_{AA}t/2}\right],
\end{equation}
i.e, $C^{13}_+(t)$ $\!\sim$ $\!g_+/\Gamma^{32}_{AA}$.
Thus, though one can allow for 
$g_+/\Gamma_{AA}^{32}$ $\!\ll$ $\!1$, this ratio 
has still to satisfy the inequality (\ref{e74}) such 
that there is a nonvanishing probability that 
one of the atoms can reach the state 
$|3\rangle$ from the initial state $|1\rangle$ to
jump to the state $|2\rangle$.

\section{Summary and conclusions}
\label{sec6}

We have proposed a scheme for non-conditional
preparation of two spatially well separated identical atoms
in long-living highly entangled states.
The scheme uses $\Lambda$-type atoms passing a 
resonator-like equipment of realistic, dispersing and 
absorbing macroscopic bodies which form electromagnetic field 
resonances, the heights and widths of which are determined
by the radiative and nonradiative (absorption) losses.
The lowest lying atomic state and the lower lying excited state, 
which can be the ground state and a metastable state
or two metastable states, play the role of the basis states 
of an atomic qubit. The atoms initially prepared in the
lowest lying states, are pumped by a single-excitation ``pulse'' 
of the body-assisted electromagnetic field,
thereby strongly driving the dipole-allowed transition 
between the lowest and highest lying atomic states.  
In this way, one of the two atoms -- we do not know which one -- 
can absorb the single-photonic excitation, and subsequent 
irreversible spontaneous decay of the excited atomic state to the 
lower lying excited state, the transition of which to the 
lowest lying state is dipole-forbidden, 
results in a metastable two-atom entangled state.

To be quite general, we have first developed the theory, without
specifying the atoms and the equipment whose body-assisted 
electromagnetic field is used for the the collective 
atom-field interaction. For the case of two $\Lambda$-type atoms,
we have derived the general solution of the coupled field-atom
evolution equations and presented special coupling conditions under
which high-degree entanglement can be achieved.
We have then applied the theory to the problem of
entanglement of two $\Lambda$-type atoms near a microsphere.
In particular, we have shown that the scheme is   
capable of realizing strong coupling in one arm and  weak coupling 
in the other arm of the $\Lambda$ configuration.
In this context, we have also analyzed the preparation of
the initial single-photonic field excitation
required for initiating the process of entanglement.        

In contrary to the common sense that the existence of 
dissipation spoils the quantum  coherence
of a system,  dissipation is here
essential to transfer the entanglement from the
strongly driven transitions to the dipole-forbidden
transitions. The fact that only ground or metastable states
serve as basis states of the qubits guarantees the 
long lifetime  of the entangled state.
It is worth noting that the scheme renders it possible
to test nonlocality  for a two-atom system.
An atomic pair  passing by a microsphere
and being entangled there, can be separated from each other 
and one can be sure that in the meantime the entanglement is not lost.

\begin{acknowledgments}
This work was supported by the Deutsche Forschungsgemeinschaft.
\"O.\c{C} and H.T.D., respectively, acknowledge support 
from the Scientific and Technical Research Council of Turkey
and the Vietnam National Program for Basic Research.
\end{acknowledgments}


\end{document}